\documentclass[journal]{IEEEtran}
\usepackage{graphicx}
\usepackage{textcomp}

\begin{document}
\title{Quality Assurance on a custom \\SiPMs array for the Mu2e
  experiment}
%
%

\author{\IEEEauthorblockN{ N. Atanov, V. Baranov, J. Budagov,  Yu. I. Davydov, V. Glagolev, V. Tereshchenko, 
Z. Usubov $                   $ }\\
\IEEEauthorblockA{Joint Institute for Nuclear Research, Dubna, Russia} \\
\and
\IEEEauthorblockN{F. Cervelli,  S. Di Falco,  S. Donati, L. Morescalchi,  E. Pedreschi, G. Pezzullo$^{*}$,  F. Raffaelli,   F. Spinella}\\
\IEEEauthorblockA{INFN sezione di Pisa, Pisa, Italy \\($*$) email: pezzullo@PI.INFN.IT}\\
\and
\IEEEauthorblockN{ $          $ F. Colao, M. Cordelli, G. Corradi, E. Diociaiuti,  R. Donghia, S. Giovannella, F. Happacher, M. Martini, \\ S. Miscetti, M. Ricci, A. Saputi, I. Sarra}\\
\IEEEauthorblockA{ Laboratori Nazionali di Frascati dell' INFN, Frascati, Italy}\\ 
\IEEEauthorblockN{ B. Echenard, D. G. Hitlin, T. Miyashita, F. Porter, R. Y. Zhu} \\
\IEEEauthorblockA{ California Institute of Technology, Pasadena, USA} \\
\IEEEauthorblockN{ F. Grancagnolo, G. Tassielli}
\IEEEauthorblockA{ INFN sezione di Lecce, Lecce, Italy }\\
\IEEEauthorblockN{ P. Murat}\\
\IEEEauthorblockA{ Fermi National Accelerator Laboratory, Batavia, Illinois, USA }\\}

\maketitle
\pagestyle{empty}
\thispagestyle{empty}

\begin{abstract}
The Mu2e experiment at Fermilab will search for the 
coherent $\mu \rightarrow e$ conversion on aluminum atoms. 
The detector system consists of a straw tube tracker and a crystal calorimeter.
A pre-production of 150 Silicon Photomultiplier arrays for the Mu2e
calorimeter has been procured. A detailed quality assurance has been
carried out on each SiPM for the determination of its own operation
voltage, gain, dark current and PDE. The measurement of the
mean-time-to-failure for a small random sample of the pro-production
group has been also completed as well as the determination of the dark
current increase as a function of the ioninizing and non-ioninizing
dose.
\end{abstract}

\begin{IEEEkeywords}
  High energy physics instrumentation, Radiation effects, Silicon radiation detectors, Nuclear physics
\end{IEEEkeywords}

\section{Introduction}
%
%
%
%
\IEEEPARstart{T}{he} Mu2e experiment at Fermilab will search for the
charged lepton flavor violating process of neutrino-less $\mu \to e$
coherent conversion in the field of an aluminum
nucleus~\cite{MU2ETDR}. Mu2e will reach a single event sensitivity of
about $2.5\cdot 10^{-17}$ that corresponds to four orders of magnitude
improvements with respect to the current best limit. The detector
system consists of a straw tube tracker and a crystal calorimeter. The
calorimeter was designed to be operable in a harsh environment where
about 10 krad/year will be delivered in the hottest region and work in
presence of 1 T magnetic field. The calorimeter role is to perform
$\mu$/e separation to suppress cosmic muons mimiking the signal, while
providing a high level trigger and a seeding the track search in the
tracker. In this paper we present the calorimeter design and the
latest R$\&$D results.

\section{Mu2e custom SiPM array}\label{REQUIREMENTS}
The Mu2e calorimeter is composed by two disks of 1348 un-doped
parallelepiped CsI crystals of 34$\times$34$\times$200 mm$^3$
dimension, each one readout by two large area SiPM
arrays~\cite{MU2ECALO}. We translated the calorimeter
requirements~\cite{CaloRef} in a series of technical specifications
for the SiPMs that are summarized in the following list:

\begin{enumerate}
\item a high gain, above 10$^6$, for each monolithic (6$\times$ 6)
  mm$^2$ SiPM cell;
\item a good photon detection efficiency (PDE) of above 20\% at 310 nm
  to well match the light emitted by the un-doped CsI crystals;
\item a large active area that, in combination with the PDE, could
  provide a light yield of above 20 p.e./MeV;
\item a fast rise time and a narrow signal width to improve time
  resolution and pileup rejection;
\item a Mean to Time Failure (MTTF) of O(10$^6$) hours;
\item and a good resilience to neutrons for a total fluency up to
  10$^{12}$ n-1MeV$_{\rm eq}$/cm$^2$.
\end{enumerate}

\begin{figure}[h!]
\centering
\includegraphics[width=3.5in]{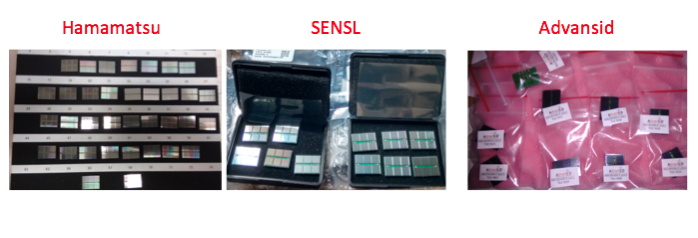}
\caption{Mu2e SiPMs from: Hamamatsu, SensL and AdvanSid.}
\label{fig:vendors}
\end{figure}

A {\em modular and custom SiPM layout} (Mu2e SiPM in the following)
has been chosen to satisfy these requirements. To well match the
wavelength of the emitted light produced by the CsI crystals, which
peaks at about 300 nm, the SiPM detection efficiency have been
extended in the UV region. The configuration readout of 2 series of
three 6$\times$6 mm$^2$ monolithic SiPMs 50$\mu$m pitch has been
selected to overcome the issues related to the parallel connection
that, due to the large capacitance, could have spoiled the pileup
rejection and the energy and time measurements.

The perforamnce of the Mu2e SiPM from three international firms, see
Figure~\ref{fig:vendors}: Hamamatsu~\cite{HAMAMATSU},
SensL~\cite{SENSL} and AdvanSid~\cite{ADVANSID} were studied to select
the vendor for the final production.

\subsection{Quality assurance procedure}
\begin{figure}[h!]
\centering
\includegraphics[width=3.5in]{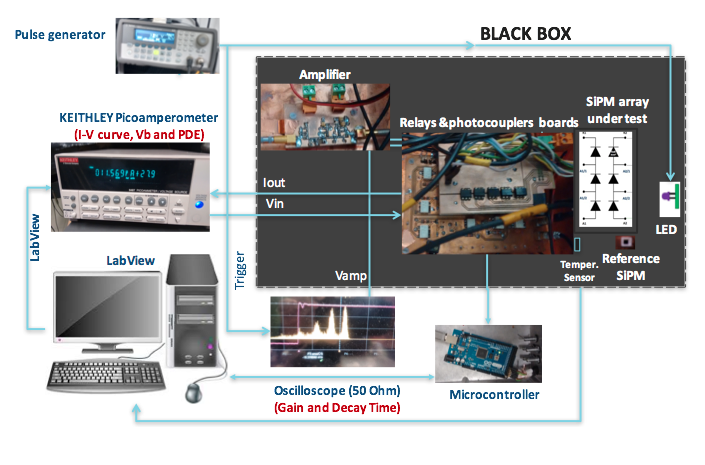}
\caption{Scheme of the semi-automatized station used for
  characterizing the Mu2e SiPMs. }
\label{fig:qa_station}
\end{figure}

A semi-automatized test station was assembled in order to measure:
gain, operational voltage, I$_{\rm dark}$ and PDE of each cell of the
Mu2e SiPMs in order to provide also information about the homogeneity
of the SiPM response. The station consisted of:
\begin{itemize}
\item source meter Keithley 6487~\cite{KEITHLEY};
\item pulse generator Agilent~\cite{AGILENT};
\item microcontroller Arduino MEGA~\cite{ARDUINO};
\item scope LeCroy WaveRunner~\cite{LECROY};
\item LED @ 315 nm from Thorlabs~\cite{LED};
\item custom relay board;
\item custom amplifier using 2 MAR8~\cite{MAR8} chips;
\item SiPM $3 \times 3$ mm$^2$ 50 $\mu$m pixel pitch~\cite{SIPMREF}.
\end{itemize}
A LabView~\cite{LABVIEW} executable was used to control the source
meter and the boards (via the
microcontroller). Figure~\ref{fig:qa_station} shows a scheme of the
setup. A water chiller system was used to keep the SiPM temperature
stable at 20\textdegree{}C.

\begin{figure}[h!]
\centering
\includegraphics[width=3.5in]{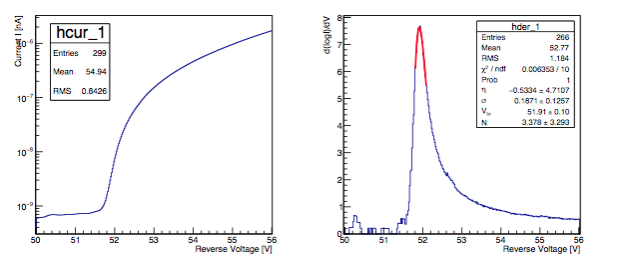}
\caption{Left: I-V$_{\rm bis}$ scan. Right: distribution of $d\log
  {\rm I_{dark}}/dV$ with included a Fit to the curve for evaluating
  the  V$_{\rm br}$. }
\label{fig:iv_curve}
\end{figure}

For each cell of the Mu2e SiPMs we measured I$_{\rm dark}$ as a
function of the bias voltage applied. This measurement allows to
evaluate the breakdown voltage V$_{\rm br}$, which corresponds to the
peak position of the $d\log {\rm I_{dark}}/dV$, and consequently the
value of the I$_{\rm dark}$ at the operation voltage V$_{\rm op}$ that
we set at V$_{\rm br}$. Figure~\ref{fig:iv_curve} shows an example of
I-V$_{\rm bis}$ scan with its logarithmic derivative.

\begin{figure}[h!]
\centering
\includegraphics[width=3.5in]{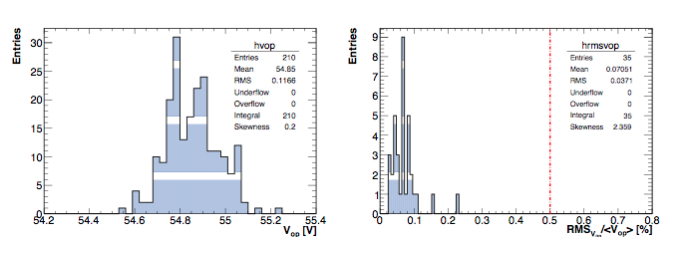}
\caption{Left: distribution of the V$_{\rm op}$ of all the Hamamatsu
  SiPM-cells. Right: of the relative V$_{\rm op}$ Root-Mean-Square
  within each Mu2e SiPM. The red line on the same plot show the 0.5\%
  threshold we required.}
\label{fig:vop}
\end{figure}

\begin{figure}[h!]
\centering
\includegraphics[width=3.5in]{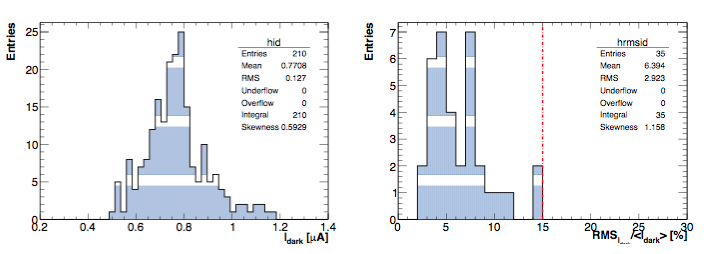}
\caption{Left: distribution of the I$_{\rm dark}$ of all the Hamamatsu
  SiPM-cells. Right: of the relative I$_{\rm dark}$ Root-Mean-Square
  within each Mu2e SiPM. The red line on the same plot show the 15\%
  threshold we required.}
\label{fig:i_dark}
\end{figure}

Figure~\ref{fig:vop} shows on the left the distribution of the V$_{\rm
  op}$ of the Hamamatsu Mu2e SiPM-cells, while the plot on the right
shows the distribution of the relative V$_{\rm op}$ Root-Mean-Square
within each Mu2e SiPM. Figure~\ref{fig:i_dark} shows on the left the
distribution of the I$_{\rm dark}$ of the same SiPM-cells and on the
right the distribution of the I$_{\rm dark}$ Root-Mean-Square within
each Mu2e SiPM under test. These measurements shows that the Hamamatsu
Mu2e SiPMs match the requirements on I$_{\rm dark}$ and V$_{\rm op}$
we specified on Section~\ref{REQUIREMENTS}.

The gain was measured at V$_{\rm op}$ using the technique described on
reference~\cite{GAINREF}; the LED was set, using the waveform
generator, in a condition where it was emitting a small amount of
light (the mean number of emitted photons detected by the SiPM cell
was $\sim$ 1), then the amplified SiPM signal was integrated in a
fixed gate of 150 ns near the peak. The resulting charge distribution
was finally use to evaluate the SiPM cell. 
\begin{figure}[h!]
\centering
\includegraphics[width=3.5in]{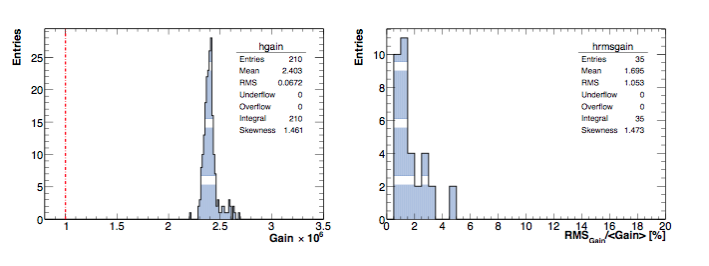}
\caption{Left: distribution of the gain of all the Hamamatsu
  SiPM-cells.The red line on the same plot show the $10^6$ threshold
  we required. Right: of the relative gain Root-Mean-Square within
  each Mu2e SiPM. }
\label{fig:gain}
\end{figure}
Figure~\ref{fig:gain} shows on the left the distribution of the
measured gain of all the SiPM-cells of the Hamamatsu Mu2e SiPMs, while
on the right the gain Root-Mean-Square within each Mu2e SiPM. The red
line on the left plot show the gain=$10^6$ threshold we required.

\begin{figure}[h!]
\centering
\includegraphics[width=3.5in]{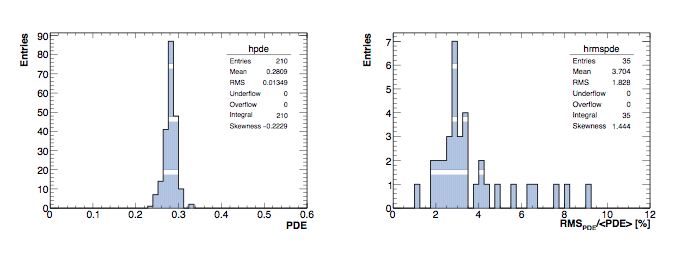}
\caption{Left: distribution of the PDE of all the Hamamatsu
  SiPM-cells. Right: of the relative PDE Root-Mean-Square within
  each Mu2e SiPM. }
\label{fig:pde}
\end{figure}

The PDE was measured by lighting the Mu2e SiPM under test and the SiPM
used as reference with the 315 nm LED and then comparing the induced
current in the two sensors. Figure~\ref{fig:pde} shows on the left the
distribution of the measured PDE on all the Mu2e SiPM cells from
Hamamatsu, while on the right the PDE Root-Mean-Square within each
Mu2e SiPM.

\section{Radiation hardness}
\begin{figure}[h!]
\centering
\includegraphics[width=3.5in]{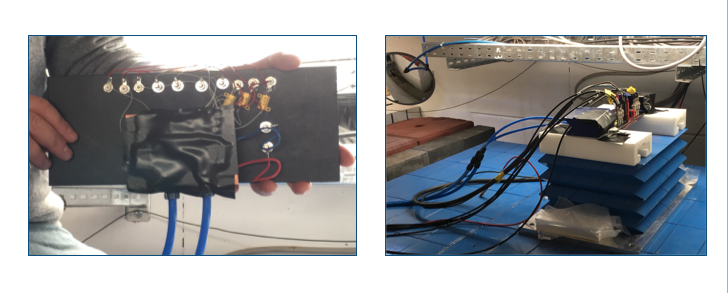}
\caption{Experimental setup used for the irradiation test.}
\label{fig:setup_neutron}
\end{figure}

\begin{figure}[h!]
\centering
\includegraphics[width=3.5in]{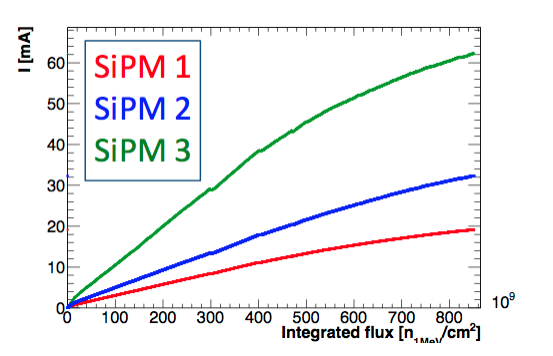}
\caption{I$_{\rm dark}$ versus the integrated neutron flux for the
  three vendors: red line is the Hamamatsu sample, blue line refers to
  the SensL one and the green line to the SiPM from AdvanSid. }
\label{fig:damage}
\end{figure}

One sample from each vendor was exposed to neutron fluence up to
10$^{12}$ n-1MeV$_{\rm eq}$/cm$^2$ at the Helmholtz-Zentrum
Dresden-Rossendorf
facility~\cite{HZDR}. Figure~\ref{fig:setup_neutron} shows the
experimental setup used to test the SiPM. During the whole irradiation
period the SiPM where kept at 20\textdegree{}C and biased at their
corresponding V$_{ \rm op}$.
Figure~\ref{fig:damage} shows the trend of I$_{\rm dark}$ as a
function of the integrated neutron flux. Same plot shows that the
Hamamatsu SiPM (red line) is more rad-hard w.r.t. the other two
vendors.

\section{Mean time to failure}
\begin{figure}[h!]
\centering
\includegraphics[width=3.5in]{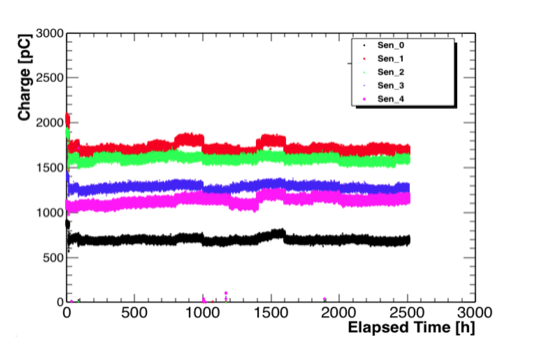}
\caption{Reconstructed charge, given in pC, of the 5 SiPMs form
  Hamamatsu under test versus the elapsed time during the MTTF test. }
\label{fig:mttf}
\end{figure}

Five samples from each vendor where kept in a custom made thermostatic
box operating at 50\textdegree{}C at their corresponding V$_{ \rm op}$
(evaluated at the same temperature) for about 2556 hours to check that
the SiPMs provide a MTTF of O(10$^6$). We daily measured the I$_{\rm
  dark}$ and the charge response, pulsing UV light with a LED on the
SiPMs, for all the samples during the entire period of
test. Figure~\ref{fig:mttf} shows the measured charge along the data
taking period for the samples from Hamamatsu. Same plot shows that all
the SiPMs were operative up to the end of the test.

\section{Summary}
We showed the preliminary results of the SiPMs pre-production for the
Mu2e calorimeter; 150 SiPM arrays where fully characterized with a
semi-automatized station and operability of the devices was also test
under neutron fluence up to 10$^{12}$ n-1MeV$_{\rm eq}$/cm$^2$. We
also verified that the MTTF of these SiPM is larger than O(10$^6$),
thus satisfying the Mu2e technical requirements.

\vfill


\section*{Acknowledgment}
This work was supported by the US Department of Energy; the Italian
Istituto Nazionale di Fisica Nucleare; the US National Science
Foundation; the Ministry of Education and Science of the Russian
Federation; the Thousand Talents Plan of China; the Helmholtz
Association of Germany; and the EU Horizon 2020 Research and
Innovation Program under the Marie Sklodowska-Curie Grant Agreement
No.690385.

%

\end{document}